\documentclass[twocolumn,secnumarabic,amssymb, amsmath, nofootinbib,tightenlines, nobibnotes, color, aps, prl,epsfig]{revtex4}
\usepackage{graphicx}% Include figure files
\usepackage{dcolumn}% Align table columns on decimal point
\usepackage{bm}% bold math
\begin{document}
\preprint{APS/123-QED}
\title{ One-to-One correspondence of soft and hard Pomeron with the CDP of the gluon density at
low $x$
}% Force line breaks with \\

\author{G.R.Boroun}%
 \email{boroun@razi.ac.ir }
\affiliation{ Department of Physics, Razi University, Kermanshah
67149, Iran}% \textbackslash\textbackslash

%\author{D.Schildknecht}%
%\email{schild@physik.uni-bielefeld.de } \affiliation{
%Fakult$\ddot{a}$t f$\ddot{u}$r Physik, Universit$\ddot{a}$t
%Bielefeld,
%D-33501 Bielefeld, Germany}% \textbackslash\textbackslash.
%\author{Masaaki Kuroda}%
%\email{kurodam@law.meijigakuin.ac.jp} \affiliation{
%***, Japan}% \textbackslash\textbackslash
\date{\today}% It is always \today, today,
             %  but any date may be explicitly specified
\begin{abstract}
The correspondence between the gluon density behavior of the color
dipole picture and the two-Pomeron approach at low $x$ deep
inelastic scattering is considered. For photon virtualities of
$Q^{2}{\gtrsim}10~\mathrm{GeV}^{2}$, the  results for the
parametrization and CDP models are defined by the CDP asymptotic
limit and are compatible with the soft and hard-Pomeron approach.
These results show that the hard-Pomeron trajectory does not
guarantee converging towards the asymptotic representation at low
and large $Q^{2}$ ($Q^{2}{\lesssim}100~\mathrm{GeV}^{2}$) values
in a wide range of the virtual-photon-proton energy squared. The
gluon distributions can be obtained directly in terms of the
 proton structure functions and the running coupling
and compared with the results from the GJR and MSTW
parametrizations.\\
%%%%%%%%%%%%%%%%%%%%%%%%%%%%%%%%%%%%%%%%%%%%%%%%%%%%%%%
\end{abstract}
 \pacs{***}%PACS, the Physics and Astronomy
                              %Classification Scheme.
\keywords{****} %Use showkeys class option if keyword
                              %display desired
\maketitle
%**********************************************************
%%%%%%%%%%%%%%%%%%%%%%%%%%%%%%%%%%%%%%%%%%%%%%%%%%%%%%%%%%%%%%%%%%%%%%%%%%%%%%%%%%%%%%%%%%%%%
\subsection{1. Introduction}
Sakurai and Schildknecht in 1972 [1] starting points on the color
dipole picture (CDP) which provides a convenient description of
deep inelastic scattering (DIS) at low $x$ [2-5]. In this picture,
the DIS cross section is factorized into a light-cone wave
function, as the virtual photon fluctuates into the
$q\overline{q}$ pair. This split is defined as a convolution of
the infinite momentum frame wave function with the pQCD calculable
coefficient functions. Then the $q\overline{q}$ pair interacts
with the gluon field in the nucleon as a gauge-invariant
color-dipole interaction. In this interaction the $q\overline{q}$
lifetime is about $1/x$ times longer than time of interaction
photon with proton [6]. At low values of the Bjorken variable
$x{\equiv}x_{\mathrm{Bj}}\simeq Q^{2}/W^{2}$, DIS of electrons on
protons defined by the processes splitting of the photon into
on-shell quark-antiquark states and interaction these states on
the proton. Here $Q^{2}$ refers to the photon virtuality and $W$
to the photon-proton center-of-mass energy. In the transverse
position space, the photoabsorption cross section at low values of
$x$ is defined by the square of the photon wave function and the
dipole cross section as [2,5,7]
\begin{eqnarray}
\sigma_{\gamma^{*}_{L,T}}(W^{2},Q^{2})&=&\int dz \int
d^{2}\mathbf{r}_{\bot}
|\Psi_{L,T}(\mathbf{r}_{\bot},z(1-z),Q^{2})|^{2}\nonumber\\
&&{\times}{\sigma}_{(q\overline{q})p}(\mathbf{r}_{\bot},z(1-z),W^{2}).
\end{eqnarray}
The variable $z$ determines the direction of the three-momentum of
the quark due to the photon direction and $|\Psi_{L,T}|^{2}$
describes the probability of the occurrence of a quark-antiquark
fluctuation with respect to the polarization direction. The dipole
cross section, related to the imaginary part of the
$(q\overline{q})p$ forward scattering amplitude. The dipole cross
section depends on the center-of-mass energy, $W$, of the
$(q\overline{q})p$ scattering process, as this implies that the
structure function reads
\begin{eqnarray}
F_{2}(x,Q^{2}){\simeq}\frac{Q^{2}}{4\pi^{2}\alpha_{EM}}
[\sigma_{\gamma^{*}_{L}p}(W^{2},Q^{2})
+\sigma_{\gamma^{*}_{T}p}(W^{2},Q^{2})].
\end{eqnarray}
Indeed, the structure function becomes a function of the single
variable $W^{2}$. In Refs.[5,7] we observe that the experimental
data corresponding to $1/W^{2}{\leq}10^{-3}$ lie on a single line
in the relevant range of $x<0.1$.\\
A new parameterization of the proton structure function which
describes fairly well the available experimental data on the
reduced cross section at low values of $x$ described in Ref.[8,9].
In Ref.[8] authors use an updated version of the global
parameterization of the ZEUS data [10] for the proton structure
function made by authors in Ref.[11]. Authors modified the fit in
two important aspects of the sieve algorithm [12] which minimizes
the squared Lorentzian  [9,11] and obtained free parameters
according to the renormalized $\chi^{2}_{min}$ per degree of
freedom whose contribution was 193.19. Analytical expression
successfully reproduces the known experimental data for
$F_{2}(x,Q^{2})$ in a wide range of $Q^{2}$ values,
$0.11{\leq}Q^{2}{\leq}1200~\mathrm{GeV}^{2}$, and Bjorken $x$
range, $10^{-6}{\leq}x{\leq}0.0494$. Then the authors in Ref.[9]
completed the parameterization method to fit all of the HERA DIS
data on $F_{2}(x,Q^{2})$ at low values of $x$ that satisfies a
saturated Froissart bound behavior. The parameterization of the
proton structure function [9] is in full accordance with the
Froissart predictions on the available experimental data in a
range of the kinematical variables $x$ and
$Q^{2}$, $x\leq 0.1$ and $0.15<Q^{2}<3000~\mathrm{GeV}^{2}$.\\
This paper is organized as follows. In the next section, the
 theoretical formalism is presented, including the parameterization of $F_{2}(x,Q^{2})$
  and the  color dipole picture.  In section 3, we present
  a detailed numerical analysis and our main results for the CDP, parametrization and Regge models. In the last
section  we summarize our main conclusions and remarks.\\
%%%%%%%%%%%%%%%%%%%%%%%%%%%%%%%%%%%%%%%%%%%%%%%%%%%%%%%%%%%%%%%%%%

\subsection{2. Theoretical formalism}

In view of low values of $x{\cong}Q^{2}/W^{2}<0.1$, the transverse
$F_{2}(x,Q^{2})$ and longitudinal $F_{L}(x,Q^{2})$ structure
functions are expressed via the gluon distribution $xg(x,Q^{2})~
(k = 2, L)$ as the approximation relation reads
\begin{eqnarray}
F_{k}(x,Q^{2}){\simeq}<e^{2}>C_{k,g}(x,Q^{2}){\otimes}xg(x,Q^{2}).
\end{eqnarray}
The symbol ${\otimes}$ denotes a convolution according to the
usual prescription,
$f(x){\otimes}g(x)=\int_{x}^{1}\frac{dy}{y}f(y)g(\frac{x}{y})$.
Here $<e^{k}>$ is the average of the charge $e^{k}$ for the active
quark flavors, $<e^{k}>=n_{f}^{-1}\sum_{i=1}^{n_{f}}e_{i}^{k}$
with $n_{f}$ as the number of considered flavors and $C_{k,g}$ are
the common Wilson coefficient functions [13]. At leading order
(LO) approximation, the  longitudinal structure function becomes
proportional to the gluon density at a rescaled value $x/\xi_{L}$
[14,15] as
\begin{eqnarray}
F_{L}(\xi_{L}x,Q^{2})=\frac{\alpha_{s}(Q^{2})}{3\pi}\sum{e_{q}^{2}xg(x,Q^{2})},
\end{eqnarray}
where the rescaling factor in the above equation has the preferred
values of $\xi_{L}{\cong}0.40$ and $\alpha_{s}(Q^{2})$ is the
running coupling at the leading order (LO, n=0) and the next-to
leading order (NLO, n=1) approximations by the following form,
respectively
\begin{eqnarray}
\alpha_{s}(Q^{2})=\frac{1}{bt}[1-n\frac{b'\ln{t}}{bt}],
\end{eqnarray}
with $b=\frac{33-2n_{f}}{12\pi}$ and
$b'=\frac{153-19n_{f}}{2\pi(33-2n_{f})}$, where
$t=\ln(\frac{Q^{2}}{\Lambda^{2}_{QCD}})$. The running of the
coupling constant $\alpha_{s}$ is determined by the
renormalization group equation $Q^2\frac{\partial
\alpha_{s}}{\partial Q^2}=\beta(\alpha_{s})$, where the $\beta$
function in QCD has the perturbative expansion [16-17]
\begin{eqnarray}
\beta(\alpha_{s})=-\alpha_{s}\sum_{n=0}^{\infty}\beta_{n}(\frac{\alpha_{s}}{4\pi})^{(n+1)}\nonumber
\end{eqnarray}
where $\beta_{0}=4\pi{b}$ and $\beta_{1}=16\pi^2{bb'}$. In
Ref.[18], the author has discussed that the running coupling of a
generic field theory can be described through a separable
differential equation involving the corresponding
$\beta$-function. Only the first loop order can be solved
analytically in terms of well-known functions. For further loop
orders the running coupling leads to transcendental equations,
where by applying an optimal Pade approximante on the
$\beta$-function, it leads to generalizations of Lambert$^{,}$s
equation. Its solution is presented in terms of a power series, as
at low $Q^{2}$ values, the appropriate NLO transcendental
equation for $\alpha_{s}$ should be used.\\
The evolution equation of the singlet distribution function  at LO
analysis at low $x$ is given by
\begin{eqnarray}
\frac{\partial{\Sigma(x,Q^{2})}}{\partial{\ln}Q^{2}}=\frac{\alpha_{s}(Q^{2})}{2\pi}
P_{qg}(\alpha_{s},x){\otimes}xg(x,Q^{2}),
 \end{eqnarray}
where $P_{qg}(\alpha_{s},x)$ is the quark-gluon splitting
function. $\Sigma$ is the singlet density and for $n_{f}=4$ reads
\begin{eqnarray}
F_{2}(x,Q^{2})=\frac{1}{4}\sum{e_{q}^{2}}x\Sigma(x,Q^{2})=\frac{5}{18}x\Sigma(x,Q^{2}).
\end{eqnarray}
A similar relation for the derivative of $F_{2}(x,Q^{2})$ with
respect to ${\ln}Q^{2}$ at low $x$ is determined by the authors in
Refs.[19-22] as
\begin{eqnarray}
\frac{\partial{F_{2}(\xi_{2}x,Q^{2})}}{\partial{\ln}Q^{2}}=\frac{\alpha_{s}(Q^{2})}{3\pi}\sum{e_{q}^{2}xg(x,Q^{2})},
\end{eqnarray}
where the rescaling factor for $F_{2}(x,Q^{2})$ has the preferred
values of $\xi_{2}{\cong}0.50$. Combining Eqs.(4) and (8), one can
calculate the longitudinal structure function by the derivative of
the structure function at a rescaled value $\eta{x}$ as
\begin{eqnarray}
F_{L}(x,Q^{2})=\frac{\partial{F_{2}(\eta{x},Q^{2})}}{\partial{\ln}Q^{2}},
\end{eqnarray}
where $\eta=\frac{\xi_{2}}{\xi_{L}}{\simeq}1.25$. In CDP, the
ratio of structure functions is dependent on the kinematic
variable $\rho{\equiv}\rho(x,Q^{2})$ by the following form
\begin{eqnarray}
\frac{F_{L}(x,Q^{2})}{F_{2}(x,Q^{2})}=\frac{1}{1+2\rho}.
\end{eqnarray}
The parameter $\rho$ is associated with the enhanced transverse
size  of $q\overline{q}$ fluctuations originating from the
difference in the photon wave functions as
\begin{eqnarray}
\frac{\sigma_{\gamma^{*}_{L}}(W^{2},Q^{2})}{\sigma_{\gamma^{*}_{T}}(W^{2},Q^{2})}
=\frac{1}{2\rho}.
\end{eqnarray}
With imposing consistency between the CDP and the pQCD, the
authors in Refs.[5,7,23] obtained the gluon distribution function
by expressing the longitudinal structure function in terms of
$F_{2}(x, Q^{2})$ as [7]
\begin{eqnarray}
\alpha_{s}(Q^{2})xg(x,Q^{2})=\frac{3\pi}{\sum{e_{q}^{2}}}\frac{1}{(2\rho+1)}
F_{2}(\xi_{L}x,Q^{2}).
\end{eqnarray}
The above equation (i.e., Eq.(12)) can be used to study the
behavior of $\alpha_{s}(Q^{2})xg(x,Q^{2})$ due to the proton
structure function.\\

$\bullet~Color~ Dipole ~Model$\\

In CDP [5,7,23], at sufficiently large $Q^{2}$ for $x<0.1$, the
structure function depends on the single variable $W^{2}$
\begin{eqnarray}
F_{2}(x,Q^{2})=F_{2}(W^{2}=\frac{Q^{2}}{x}),
\end{eqnarray}
which is consistent with the experimental data with an eye-ball
fit by the form
\begin{eqnarray}
F_{2}(W^{2})=f_{2}(\frac{W^{2}}{1\mathrm{GeV}^{2}})^{C_{2}},
\end{eqnarray}
with $f_{2}=0.063$ and $C_{2}=0.29$. In this picture, the known
expression for $\alpha_{s}(Q^{2})xg(x,Q^{2})$ (with respect to
Eqs.(12) and (14)) reads as follows
\begin{eqnarray}
\alpha_{s}(Q^{2})xg(x,Q^{2})=\frac{3\pi}{(2\rho+1)\sum{e_{q}^{2}}}\frac{f_{2}}{\xi_{L}^{C_{2}}}(\frac{W^{2}}{1\mathrm{GeV}^{2}})^{C_{2}},
\end{eqnarray}
where $\rho=4/3$ and $\sum{e_{q}^{2}}=10/9$ for four active
flavors.\\

$\bullet~Parametrization ~Model$\\

The proton structure function is parametrized with a  global fit
function [9] with a combined fit to the H1 and ZEUS Collaboration
data for $F_{ 2}(x,Q^{2})$ at $0.15< Q^{2} < 3000~
\mathrm{GeV}^{2}$ and $x< 0.01$ takes the form
\begin{eqnarray}
F_{ 2}(x,Q^{2})=D(Q^{2})(1-x)^{n}\sum_{m=0}^{2}A_{m}(
Q^{2})L^{m}(x,Q^{2}),
\end{eqnarray}
where
\begin{eqnarray}
D(Q^{2})&=&\frac{Q^{2}(Q^{2}+\lambda
M^{2})}{(Q^{2}+M^{2})^{2}},\nonumber\\
A_{0}(Q^{2})&=&a_{00}+a_{01}L_{2}(Q^{2}),\nonumber\\
A_{i}(Q^{2})&=&\sum_{k=0}^{2}a_{ik}L_{2}(Q^{2})^{k},~~i=(1,2),\nonumber\\
L(Q^{2},x)&=&\ln\frac{1}{x}+{\ln}\frac{Q^{2}}{Q^{2}+\mu^{2}},\nonumber\\
L_{2}(Q^{2})&=&{\ln}\frac{Q^{2}+\mu^{2}}{\mu^{2}},
\end{eqnarray}
and the effective parameters are defined in Table I. Note that the
results of using the parametrization method  can be found recently
in Refs.[24-26]. In this picture, the known expression for
$\alpha_{s}(Q^{2})xg(x,Q^{2})$ (with respect to Eqs.(12) and (16))
reads as follows
\begin{eqnarray}
\alpha_{s}(Q^{2})xg(x,Q^{2})&=&\frac{3\pi}{(2\rho+1)\sum{e_{q}^{2}}}D(Q^{2})(1-\xi_{L}x)^{n}\nonumber\\
&&\sum_{m=0}^{2}A_{m}( Q^{2})L^{m}(\xi_{L}x,Q^{2}).
\end{eqnarray}

$\bullet~Two-Pomeron ~Model$\\

HERA has shows that the $\sigma^{\gamma^{*}p}$ rapid rise with
$W^{2}$ as an effective power
\begin{eqnarray}
\sigma^{\gamma^{*}p}\sim F(Q^{2})(W^{2})^{\lambda(Q^{2})},
\end{eqnarray}
where the power $\lambda(Q^{2})$ has been extracted by H1
Collaboration in Ref.[27]. The effective-power behavior of the
proton structure function at low $x$ corresponds to
\begin{eqnarray}
F_{2}(x,Q^{2}) \sim f(Q^{2})x^{-\lambda(Q^{2})}.
\end{eqnarray}
Rather, the authors in Ref.[28] show that one should parametrize
the data with a sum of fixed powers of $x$ in the two-Pomeron
model as
\begin{eqnarray}
F_{2}(x,Q^{2}) &\sim & \sum _{i=0,1}
f_{i}(Q^{2})x^{-\epsilon_{i}}\nonumber\\
&&=F_{2}^{\mathrm{hard}}+F_{2}^{\mathrm{soft}},
\end{eqnarray}
where the  $i=0$ term is hard-Pomeron exchange and  $i=1$ term is
soft-Pomeron exchange. The authors in Ref.[28] showed that a very
good fit to data (for $Q^{2}=0$ to $5000~\mathrm{GeV}^{2}$) was
provided by the economical parameterization as
\begin{eqnarray}
f_{0}(Q^{2})&=&\frac{A_{0}(Q^{2})^{1+\epsilon_{0}}}{(1+Q^{2}/Q^{2}_{0})^{1+\epsilon_{0}/2}},\nonumber\\
f_{1}(Q^{2})&=&\frac{A_{1}(Q^{2})^{1+\epsilon_{1}}}{(1+Q^{2}/Q^{2}_{1})^{1+\epsilon_{1}}},
\end{eqnarray}
where $\epsilon_{0}=0.452$, $\epsilon_{1}=0.0667$,
$A_{0}=0.00151$, $A_{1}=0.658$, $Q_{0}^{2}=7.85$ and
$Q_{1}^{2}=0.658$.\\
In this model, the known expression for
$\alpha_{s}(Q^{2})xg(x,Q^{2})$ (with respect to Eqs.(12) and (21))
reads as follows
\begin{eqnarray}
\alpha_{s}(Q^{2})xg(x,Q^{2})&=&\frac{3\pi}{(2\rho+1)\sum{e_{q}^{2}}}[f_{0}(Q^{2})(\xi_{L}x)^{-\epsilon_{0}}\nonumber\\
 &&+f_{1}(Q^{2})(\xi_{L}x)^{-\epsilon_{1}}].
\end{eqnarray}

$\bullet~Tensor-Pomeron ~Model$\\

Recently, in Ref.[29], the validity of the CDP with respect to the
tensor-Pomeron model [30] for photon virtualities of
$Q^{2}{\gtrsim}20~\mathrm{GeV}^{2}$ is considered. Consistency of
the models has shown that the CDP with the perturbative QCD (pQCD)
improved parton model at low $x$. Authors in Ref.[30] applied the
tensor-Pomeron model to low-$x$ deep-inelastic lepton-nucleon
scattering and photoproduction due to the  center-of-mass energies
in the range $6-318~\mathrm{GeV}$ and
$Q^{2}{\lesssim}50~\mathrm{GeV}^{2}$. The hadronic high-energy
reactions defined with respect to the two- Pomeron-plus-Reggeon
approach in Ref.[30]. The virtual Compton amplitude in the
tensor-Pomeron approach for large $W^{2}$ is defined  by the
exchange of the two Pomerons, $\mathbb{P}{0}$ and $\mathbb{P}{1}$,
plus the $f_{2}R$ reggeon. Therefore the proton structure
function, in the tensor-Pomeron model, leads
\begin{eqnarray}
F_{2}(W^{2},Q^{2})&=&\frac{Q^{2}}{\pi}(1-x)[1+2\delta((W^{2},Q^{2}))]^{-1}\frac{W^{2}-m_{p}^{2}}{W^{2}}\nonumber\\
&&3{\sum}_{j=0,1,2}\widehat{b}_{j}(Q^{2})\beta_{jpp}(W^{2}\widetilde{\alpha}'_{j})^{\varepsilon_{j}}\cos(\frac{\pi}{2}\varepsilon_{j})\nonumber\\
&&[1+\frac{2Q^{2}}{W^{2}-m_{p}^{2}}+\frac{Q^{2}(Q^{2}+m_{p}^{2})}{(W^{2}-m_{p}^{2})^{2}}],
\end{eqnarray}
where $$
\delta(W^{2},Q^{2})=\frac{2m_{p}^{2}Q^{2}}{(W^{2}+Q^{2}-m_{p}^{2})^{2}}.$$
The parameters of the tensor-Pomeron approach with hard and soft
Pomeron and $f_{2}R$ reggeon exchange are
\begin{eqnarray}
\mathrm{Soft}:\hspace{1cm}{\alpha}'_{1}=\widetilde{\alpha}'_{1}=0.25~\mathrm{GeV}^{-2},~{\varepsilon}_{1}=\alpha_{1}(0)-1,\nonumber\\
{\varepsilon}_{1}=0.0935( ^{+76}_{-64}),\hspace{3.5cm}\nonumber\\
\mathrm{Hard}:\hspace{1cm}{\alpha}'_{0}=\widetilde{\alpha}'_{0}=0.25~\mathrm{GeV}^{-2},~{\varepsilon}_{0}=\alpha_{0}(0)-1,\nonumber\\
~\varepsilon_{0}=0.3008( ^{+73}_{-84}),\hspace{3.5cm}\nonumber\\
\mathrm{Reggeon}:~{\alpha}'_{2}=\widetilde{\alpha}'_{2}=0.90~\mathrm{GeV}^{-2},\alpha_{2}(0)=0.485(
^{+88}_{-90}),\nonumber
\end{eqnarray}
where $\beta_{0pp}=\beta_{1pp}=1.87~\mathrm{GeV}^{-1}$,
$\beta_{2pp}=3.68~\mathrm{GeV}^{-1}$ and
$m_{p}=0.938~\mathrm{GeV}$. The values of coupling functions,
$\widehat{b}_{j}(Q^{2})$, obtained in the fit HERA DIS and
photoproduction data read
\begin{eqnarray}
\widehat{b}_{0}(10~\mathrm{GeV}^{2})&=&\exp(-5.669(
^{+99}_{-101}))~\mathrm{GeV}^{-1}\nonumber\\
\widehat{b}_{0}(50~\mathrm{GeV}^{2})&=&\exp(-6.899(
^{+78}_{-80}))~\mathrm{GeV}^{-1}\nonumber\\
\widehat{b}_{1}(10~\mathrm{GeV}^{2})&=&\exp(-4.668(
70))~\mathrm{GeV}^{-1}\nonumber\\
\widehat{b}_{1}(50~\mathrm{GeV}^{2})&=&\exp(-7.870(
29))~\mathrm{GeV}^{-1}\nonumber
\end{eqnarray}
and
\begin{eqnarray}
\widehat{b}_{2}(Q^{2})&=&c_{2}\exp(-Q^{2}/d_{2})\nonumber\\
{c}_{2}&=&\exp(-0.38(
^{+36}_{-35}))~\mathrm{GeV}^{-1}\nonumber\\
{d}_{2}&=&\exp(-1.35( ^{+34}_{-35}))~\mathrm{GeV}^{-2},\nonumber
\end{eqnarray}
where the uncertainties indicated in the above brackets are
determined using the MINOS algorithm. Therefore, the gluon
distribution multiplied by the running coupling (with respect to
Eqs.(12) and (24)) reads
\begin{eqnarray}
\alpha_{s}(Q^{2})xg(x,Q^{2})&=&\frac{3\pi}{(2\rho+1)\sum{e_{q}^{2}}}\frac{Q^{2}}{\pi}(1-\xi_{L}x)\nonumber\\
&&3[1+2\delta((W^{2},Q^{2}))]^{-1}\frac{W^{2}-m_{p}^{2}}{W^{2}}\nonumber\\
&&{\sum}_{j=0,1,2}\widehat{b}_{j}(Q^{2})\beta_{jpp}(W^{2}\widetilde{\alpha}'_{j})^{\varepsilon_{j}}\cos(\frac{\pi}{2}\varepsilon_{j})\nonumber\\
&&[1+\frac{2Q^{2}}{W^{2}-m_{p}^{2}}+\frac{Q^{2}(Q^{2}+m_{p}^{2})}{(W^{2}-m_{p}^{2})^{2}}].
\end{eqnarray}
In the following, we employ the parameterization of the proton
structure function, two and tensor-Pomeron models in CDP and
consider the behavior of the $\alpha_{s}(Q^{2})xg(x,Q^{2})$ into
the center-of-mass energy $\mathrm{W}$, due to Eqs.(15, 18, 23)
and (25), in the next section.\\

\subsection{3. Results and discussions}

We have calculated the $W^{2}$-dependence, at low $x$, of
$\alpha_{s}(Q^{2})xg(x,Q^{2})$ according to the CDP and compared
with the parametrization and Pomeron models . Results of
calculations and comparisons  are presented in Figs. 1 up to 4. In
Fig.1 we compared the behavior of the gluon distribution with
respect to the CDP asymptotic limit (Eq.(15)) and the
parametrization of the proton structure function (Eq.(18)) for
$1~\mathrm{GeV}^{2}{\leq}Q^{2}{\leq}100~\mathrm{GeV}^{2}$ in a
wide range of $\mathrm{W}^{2}$. In this figure (i.e., Fig.1) we
observe that the parametrization model results converge towards
the CDP asymtotic limit for $Q^{2}{\gtrsim}10~\mathrm{GeV}^{2}$.\\
In Fig.2 we compared the results in Fig.1 with respect to the CDP
results  at any $Q^{2}$ value. These results are based on
$F_{2}(\xi_{L}x,Q^{2})$ according to (12) and are comparable with
the CDP asymptotic limit and the parametrization model at
moderate and large $Q^{2}$ values.\\
\begin{figure}[h]
\includegraphics[width=0.55\textwidth]{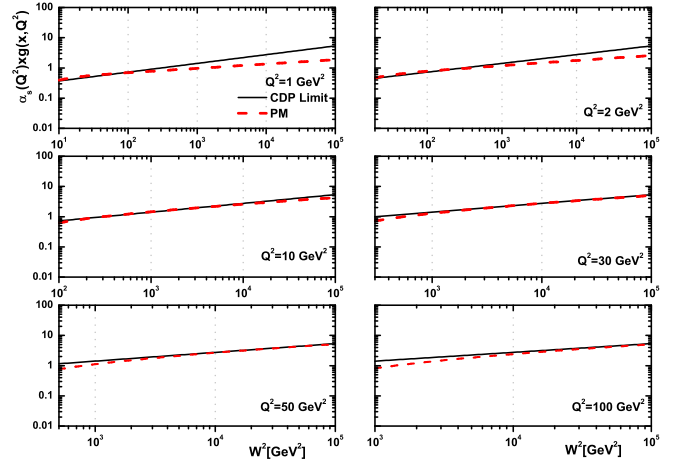}
\caption{$\alpha_{s}(Q^{2})xg(x,Q^{2})$ as a function of $W^{2}$
for various values of $Q^{2}$ in the CDP asymptotic limit
(Eq.15)(solid curve) and parametrization model (PM,Eq.18)(dashed
curve ).}\label{Fig1}
\end{figure}
\begin{figure}[h]
\includegraphics[width=0.55\textwidth]{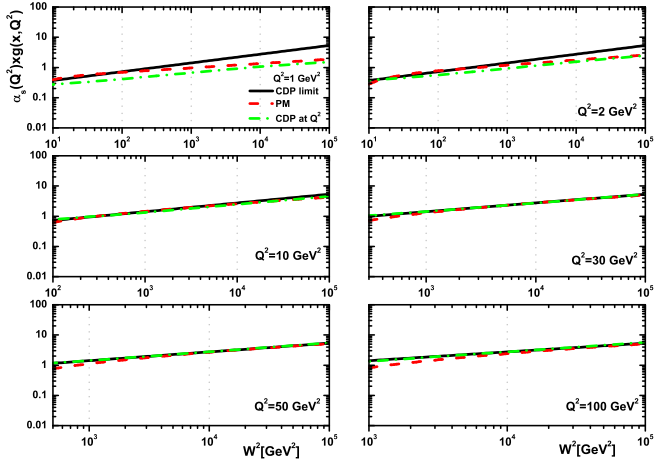}
\caption{Same as Fig.1 but based on (Eq.12)(dashed-dot curve).
}\label{Fig2}
\end{figure}
In Fig.3, we compared the Regge behavior with the parametrization
model and the CDP predictions. The parametrization model is
comparable with the Regge behavior at low and large $Q^{2}$ values
and it is comparable with CDP at moderate and high $Q^{2}$ values.
In this figure (i.e., Fig.3), the Regge behavior is defined into
the soft and hard Pomeron behaviors. The behavior of the
two-Pomeron approach converge towards the CDP and parametrization
models at $Q^{2}{\gtrsim}10~\mathrm{GeV}^{2}$.  Consistency
between results for moderate and large $Q^{2}$ values shows that
two-Pomeron approach leads to the CDP asymptotic limit where it is
free of $Q^{2}$ parameters. This indicates that the Regge model
must have at least two parameters or more to match the models. In
Ref.[30] two-Pomeron-plus-Reggeon approach fitted to the
experimental data on the deep-inelastic lepton-nucleon scattering
at low values of $x$ and consistency with the CDP is introduced in
Ref.[29].\\
\begin{figure}[h]
\includegraphics[width=0.55\textwidth]{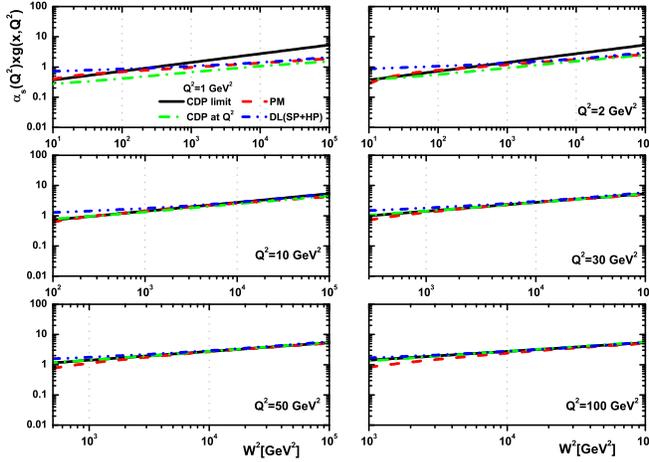}
\caption{Same as Fig.2, and compared with soft and hard Pomeron
model (SP+HP,Eq.23) (dased-dot-dot curve). }\label{Fig3}
\end{figure}
In fact, we have shown that in order to converge the CDP results
with the Regge theory, it is necessary to introduce the Regge
theory with two-Pomeron  approach.
\begin{figure}
\includegraphics[width=0.55\textwidth]{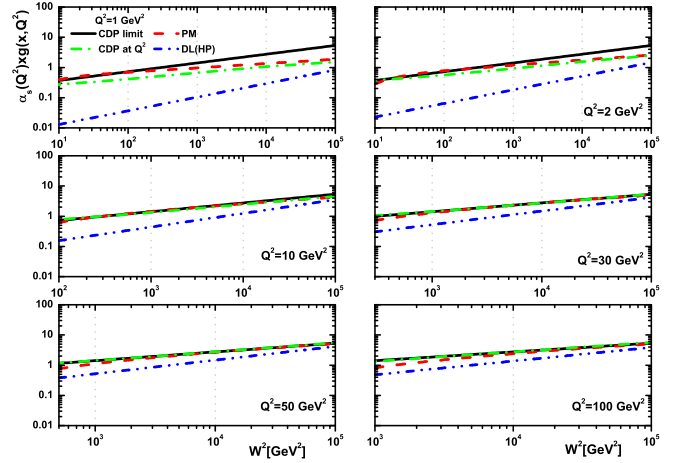}
\caption{Same as Fig.3, and compared with  hard Pomeron model
(HP,Eq.23) (dased-dot-dot curve). }\label{Fig4}
\end{figure}
In this connection, we plot $\alpha_{s}(Q^{2})xg(x,Q^{2})$
according to Eqs.(12,15) and (18) and compared the Regge theory
with respect to the only hard-Pomeron trajectory due to Eq.(23).
In Fig.4, we observe that the hard-Pomeron approach result do not
converge with the  CDP and parametrization models for
$1~\mathrm{GeV}^{2}{\leq}Q^{2}{\leq}100~\mathrm{GeV}^{2}$ in a
wide range of $\mathrm{W}^{2}$. Indeed, the results obtained by
employing the CDP representation of the experimental data of the
proton structure function in the pQCD improved parton model do not
support the necessity of modifications by single (hard) Pomeron
effects at $Q^{2}{\leq}100~\mathrm{GeV}^{2}$. We can observe that
there is  asymptotically agree between the parametrization and
hard-Pomeron results for $Q^{2}{>}100~\mathrm{GeV}^{2}$. In
summary  two Pomeron models (soft+hard) improve results in
comparison with  the CDP and parametrization models  at
$1~\mathrm{GeV}^{2}{\leq}Q^{2}{\leq}100~\mathrm{GeV}^{2}$. Soft
and hard Pomeron models already discussed in Ref.[31]. The
inclusive electroproduction on the proton using a soft and a hard
Pomeron have been studied in Ref.[31].\\
The derivative of the hard-Pomeron behavior from the eye-ball fit
can be improved again by considering the tensor-Pomeron behavior
of the proton structure function in Fig.4. In Ref.[30] the soft
and hard Pomeron coupling functions are obtained for $Q^{2}=10$
and $50~\mathrm{GeV}^{2}$. Therefore in Fig.5 we show that results
obtained in Fig.4 are comparable with others as we used the
tensor-Pomeron approach of the proton structure function.
According to (25), the gluon distribution may equivalently be
expressed in terms of the structure function due to the
tensor-Pomeron approach. As seen in Fig.5, the results at
$Q^{2}=10$ and $50~\mathrm{GeV}^{2}$ are comparable with others.
The results obtained by employing the tensor-Pomeron approach are
comparable with CDP representation
than the hard-Pomeron approach.\\
\begin{figure}
\includegraphics[width=0.55\textwidth]{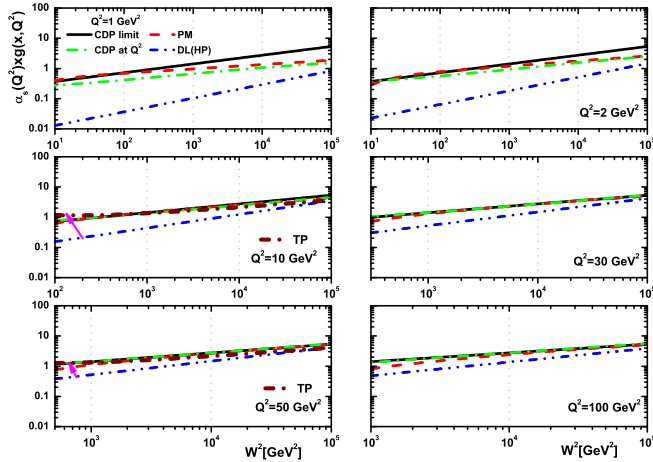}
\caption{Same as Fig.4. The tensor-Pomeron (TP) behavior of the
proton structure function plotted at $Q^{2}=10$ and
$50~\mathrm{GeV}^{2}$ (TP,Eq.25) (short-dash-dot curve).
}\label{Fig4}
\end{figure}
\begin{figure}
\includegraphics[width=0.45\textwidth]{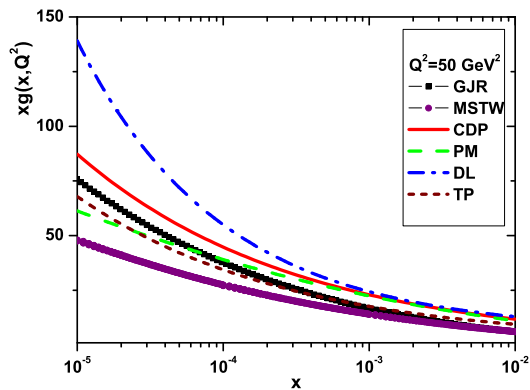}
\caption{The gluon distributions for $Q^{2}=50~\mathrm{GeV}^{2}$
by using the CDP [5,7,23], PM [9], DL [28] and TP [30] models
compared to the GJR [33] and MSTW [34] parameterizations.
}\label{Fig4}
\end{figure}
To conclude the low-$x$ analysis one needs the explicit
expressions for the gluon distribution, $xg(x,Q^2)$, in Eq.(12) as
\begin{eqnarray}
xg(x,Q^{2})=\frac{3\pi
}{\alpha_{s}(Q^{2})\sum{e_{q}^{2}}}\frac{1}{(2\rho+1)}
F_{2}(\xi_{L}x,Q^{2}),
\end{eqnarray}
where the proton structure functions are defined in Eqs.(14),
(16), (21) and (24) according to the CDP [5,7,23], PM [9], DL [28]
and TP [30] models respectively. With the explicit form of the
standard representation for QCD couplings at the LO and NLO
approximations in Eq.(5), the behavior of gluon distribution is
investigated. In Ref.[32] the authors defined the renormalization
group equations of the QCD running coupling and quark masses in a
mathematically strict way. The QCD scale parameter $\Lambda$ has
been extracted from the running coupling $\alpha_{s}$ normalized
at the $Z$-boson mass, $\alpha_{s}(M_{Z}^{2})$, where
$\Lambda^{n_{f}=4}_{\mathrm{LO}}=119~\mathrm{MeV}$ and
$\Lambda^{n_{f}=4}_{\mathrm{NLO}}=322~\mathrm{MeV}$. We have
calculated the $x$-dependence of the gluon distribution function
as described above in the NLO approximation. In Fig.6, we present
the comparison of the gluon distribution from the CDP [5,7,23], PM
[9], DL [28] and TP [30] models, with the results based on the GJR
[33] and MSTW [34] parameterizations at $
Q^{2}=50~\mathrm{GeV}^{2}$. As can be seen, the values of the
gluon distribution function increase as $x$ decreases. This figure
indicates that the obtained results from the present analysis,
based on the CDP are compatible with the ones obtained from the
parametrization methods.\\

\subsection{4. Conclusion}

In conclusion, we have studied the effects of soft and hard
Pomeron in relation to the CDP and Parametrization models. We
determined the gluon distribution function (multiplied by
$\alpha_{s}(Q^{2})$) from our representation of the
photoabsorption cross section in the CDP and compared with the
results of the parametrization of the proton structure function
and the soft and hard Pomeron in the structure function of the
proton. It turned out that the gluon function at order
$\alpha_{s}(Q^{2})$ is proportional to the  proton structure
function  at a shifted scale $x{\rightarrow}\xi_{L}x$. The
parametrization and CDP results have similar behavior at
$10~\mathrm{GeV}^{2}{\leq}Q^{2}{\leq}100~\mathrm{GeV}^{2}$. We
have found that the Regge like behavior of the proton structure
function with a soft and hard Pomeron and also tensor-Pomeron
approach improve the description of the gluon behavior at
$10~\mathrm{GeV}^{2}{\leq}Q^{2}{\leq}100~\mathrm{GeV}^{2}$. The
soft and hard Pomeron results confirm the predictions of the CDP
and this is requiring consistency of the CDP and pQCD for
$Q^{2}{\gtrsim}10~\mathrm{GeV}^{2}$.\\
Furthermore, we have shown that the gluon distribution from the
proton structure function can be estimated with respect to the
running coupling in a general model using the CDP model. Explicit,
analytical expressions for the gluon distribution function are
obtained in terms of the effective parameters of the proton
structure function in CDP, PM, DL and TP models and results of
numerical calculations as well as comparisons with GJR and MSTW
parametrizations are presented.\\
%%%%%%%%%%%%%%%%%%%%%%%%%%%%%%%%%%%%%%%%%%%%%%%%%%%%%%

\subsection{ACKNOWLEDGMENTS}

The author is grateful to Razi University for the financial
 support of this project.\\

%%%%%%%%%%%%%%%%%%%%%%%%%%%%%%%%%%%%%%%%%%%%%%%%%%%%%%%%%%%%%%%%%%%%%%%%
\begin{table}[h]
\caption{ The effective parameters at low $x$ are defined by the
Block-Halzen fit to the real photon-proton cross section as
$M^{2}=0.753 \pm 0.068~ \mathrm{GeV}^{2}$, $\mu^2 = 2.82 \pm
0.290~ \mathrm{GeV}^{2}$, $n=11.49\pm 0.99$ and $\lambda=
2.430~\pm 0.153$  [9].}
\begin{tabular} {cccc}
\toprule \\  \multicolumn{2}{c}{parameters \quad \quad \quad ~~~~~~~~~~~~~~~~value}    \\ &&&\\ \hline \\ &&&\\
  $a_{10} $  &   \quad  $8.205\times 10^{-4}~~  \pm  4.62\times10^{-4} $  \\

  $a_{11} $  &   \quad   $-5.148\times 10^{-2}\pm 8.19\times10^{-3}$  \\

  $a_{12}$   &    \quad  $-4.725\times 10^{-3}\pm 1.01\times10^{-3}$   \\  &&&\\

 $a_{20}$   &   \quad   $2.217\times 10^{-3}\pm 1.42\times10^{-4} $ \\

 $a_{21}$   &   \quad   $1.244\times 10^{-2}\pm 8.56\times10^{-4}$  \\

 $a_{22}$    &    \quad  $5.958\times 10^{-4}\pm 2.32\times10^{-4} $ \\ &&& \\

$a_{00}$& \quad  $2.550\times 10^{-1}~\pm 1.600\times10^{-2}$ & &\\

$a_{01}$& \quad  $1.475\times 10^{-1}~\pm 3.025\times10^{-2}$ & &\\

\hline

\end{tabular}
\end{table}

%%%%%%%%%%%%%%%%%%%%%%%%%%%%%%%%%%%%%%%%%%%%%%%%%%%
\section{References}
1. J.J.Sakurai and D.Schildknecht, Phys.Lett.B{\bf40}, 121(1972);
B.Gorczyca and D.Schildknecht, Phys.Lett.B{\bf47},
71(1973).\\
2. N.N.Nikolaev and B.G.Zakharov, Z.Phys.C{\bf49}, 607(1991);
N.N.Nikolaev and B.G.Zakharov, Z.Phys.C{\bf53}, 331(1992).\\
3. K.Golec-Biernat and M.W$\mathrm{\ddot{u}}$sthoff,
Phys.Rev.D{\bf59}, 014017(1998); H.Kowalski, L.Motyka and G.Watt,
Phys.Rev.D{\bf74}, 074016(2006).\\
4. M.R.Pelicer et al., Eur.Phys.J.C{\bf79}, 9(2019); B.Sambasivam,
T.Toll and T.Ullrich;
 Phys.Lett.B{\bf803}, 135277(2020); G.M.Peccini, F.Kopp, M.V.T.Machado and D.A.Fagundes,
 Phys.Rev.D{\bf101}, 074042 (2020).\\
5. M.Kuroda and D.Schildknecht, Phys.Lett. B{\bf618}, 84(2005);
 M.Kuroda and D.Schildknecht, Phys.Rev. D{\bf96}, 094013(2017);
 D.Schildknecht, B.Surrow and M.Tentynkov,
 Eur.Phys.J.C{\bf20}, 77(2001); M.Kuroda and D.Schildknecht, International Journal of Modern Physics A{\bf31}, No. 30, 1650157 (2016);
  G.Cvetic, D.Schildknecht and A.Shoshi, Eur.Phys.J.C{\bf13},
301(2000); D.Schildknecht, Nuclear Physics B Proceedings
Supplement{\bf00},1 (2012);\\
6. C.Ewerz, A.von Manteuffel and O.Nachtmann, JHEP{\bf03},
102(2010); C.Ewerz, A. von Manteuffel
 and O.Nachtmann, Phys.Rev.D{\bf77}, 074022(2008).\\
7.  M.Kuroda and D.Schildknecht,
Phys.Rev.D{\bf85}, 094001(2011).\\
8.  M. M. Block and L. Durand,  arXiv [hep-ph]: 0902.0372
(2009).\\
9. M. M. Block, L. Durand and P. Ha, Phys. Rev.D{\bf 89}, 094027 (2014).\\
10. J.Breitweg et al. (ZEUS), Phys.Lett.B{\bf487}, 53(2000);
S.Chekanov et al. (ZEUS), Eur.Phys.J.C{\bf21}, 443(2001).\\
11. E.L.Berger, M.M.Block and C.-I.Tan, Phys.Rev.Lett.{\bf98},
242001(2007).\\
12. M.M.Block, Nucl.Inst. and Meth.A.{\bf556}, 308(2006).\\
13. R.G.Roberts, The structure of the proton:Deep
inelastic scattering, Cambridge University Press (1990).\\
14. A.D. Martin et al., Phys. Rev. D{\bf37}, 1161(1988); A.M.
Cooper-Sarkar et al., Z. Phys. C{\bf39}, 281(1988); A.M.
Cooper-Sarkar et al., Acta Phys.Pol.B{\bf34}, 2911(2003).\\
15. G.R.Boroun and B.Rezaei, Eur.Phys.J.C{\bf72}, 2221(2012).\\
16.  H.D.Politzer, Phys.Rev.Lett. {\bf30}, 1346 (1973); D.J Gross
and F.Wilczek, Phys.Rev.Lett. {\bf30}, 1343 (1973); W.A.Caswell,
Phys.Rev.Lett. {\bf33}, 244 (1974).\\
17. O.V.Tarasov, A.A.Vladimirov and A.Yu.Zharkov, Phys.Lett.B
{\bf93}, 429 (1980); S.A.Larin and J.A.M.Vermaseren, Phys.Lett.B
{\bf303}, 334 (1993); R.K.Ellis, W.J.Stirling and B.R.Webber,
$"$QCD and Collider Physics$"$, Cambridge university press,
(1996).\\
18. J.$\ddot{\mathrm{O}}$sterman, arXiv [math-ph]: 1912.08016.\\
19. L.N. Lipatov, Sov. J. Nucl. Phys.{\bf20}, 94(1975); V.N.
Gribov, L.N. Lipatov, Sov. J. Nucl. Phys.{\bf15}, 438 (1972); G.
Altarelli, G. Parisi, Nucl. Phys. B{\bf126}, 298(1977); Yu.L.
Dokshitzer, Sov. Phys.
JETP {\bf46}, 641 (1977).\\
20. K. Prytz, Phys. Lett. B {\bf311}, 286(1993).\\
21. G.R.Boroun, Phys.Rev.C{\bf97}, 015206(2018).\\
22. G.R.Boroun and B.Rezaei, Eur.Phys.J.C{\bf73}, 2412(2013).\\
23. D.Schildknecht, arXiv [hep-ph]:1104.0850 (2011).\\
24. G.R.Boroun, Eur.Phys.J.A {\bf57}, 219 (2021).\\
25. L.P.Kaptari et al., Phys.Rev.D{\bf99}, 096019 (2019).\\
26. G.R.Boroun and B.Rezaei, Phys.Rev.C {\bf103}, 065202 (2021).\\
27. C.Adloff et al.[H1 Collab.], Phys.Lett.B{\bf520}, 183
(2001).\\
28. P.V.Landshoff, arXiv [hep-ph]:0203084 (2002); A.Donnachie and
P.V.Landshoff, Phys.Lett.B{\bf595}, 393 (2004); A.Donnachie and
P.V.Landshoff, Phys.Lett.B{\bf533}, 277 (2002); A.Donnachie and
P.V.Landshoff, Phys.Lett.B{\bf550}, 160 (2002).\\
29. D.Schildknecht, Phys.Rev.D {\bf104}, 014009 (2021); G.R.Boroun, M.Kuroda and D.Schildknecht, arXiv [hep-ph]: 2206.05672.\\
30. D.Britzger, C.Ewerz, S.Glazov, O.Nachtmann and S.Schmitt,
Phys.Rev.D {\bf100}, 114007 (2019).\\
31. J.R.Cudell, A.Lengyel, E.Martynov and O.V.Selyugin,
Nucl.Phys.B Proc.Suppl. {\bf152}, 79 (2006); arXiv [hep-ph]:
0710.5391; U.D$^{,}$Alesio, A.Metz and H.J.Pirner, Eur.Phys.J.C
{\bf9},
601(1999).\\
32. H.M.Chen et al., International Journal of Modern Physics
E{\bf31}, No.02, 2250016 (2022).\\
33. M. Gluck, P. Jimenez-Delgado, E. Reya, Eur.Phys.J.C {\bf53},
355 (2008).\\
34. A. Martin, W. Stirling, R. Thorne, and G. Watt, Eur.Phys.J. C
{\bf63}, 189 (2009).\\
%%%%%%%%%%%%%%%%%%%%%%%%%%%%%%%%%%%%%%%%%%%%%%%%

%%%%%%%%%%%%%%%%%%%%%%%%%%%%%%%%%%%%%%%%%%%%%%%%%%%%%%%%
\end{document}